\documentclass[%
 reprint,
 amsmath,amssymb,
 aps,
]{revtex4-2}

\usepackage{graphicx}
\usepackage{dcolumn}
\usepackage{bm}

\begin{document}

\preprint{APS/123-QED}
\title{Proposal for measuring Newtonian constant of gravitation at an  exceptional point  in an optomechanical system}


\author{Lei Chen}
 \altaffiliation[Also at ]{Key Laboratory of Artificial Structures and Quantum Control (Ministry of Education), School of Physics and Astronomy, Shanghai Jiao Tong University, 800 Dong Chuan Road, Shanghai 200240, China}
 \email{chenleiquantum99@sjtu.edu.cn}
\author{Jian Liu}
 \altaffiliation[Also at ]{Key Laboratory of Artificial Structures and Quantum Control (Ministry of Education), School of Physics and Astronomy, Shanghai Jiao Tong University, 800 Dong Chuan Road, Shanghai 200240, China}
\author{Ka-di Zhu}%
\affiliation{%
 Key Laboratory of Artificial Structures and Quantum Control (Ministry of Education), School of Physics and Astronomy, Shanghai Jiao Tong University, 800 Dong Chuan Road, Shanghai 200240, China
}%





\begin{abstract}
We develop a quantum mechanical method of measuring the Newtonian constant of gravitation, $G$. In this method, an optomechanical system consisting of two cavities and two membrane resonators is used. The added source mass would induce the shifts of the eigenfrequencies of the supermodes. Via detecting the shifts, we can perform our measurement of $G$. Furthermore, our system can features exceptional point (EP) which are branch point singularities of the spectrum and eigenfunctions. In the paper, we demonstrate that operating the system at EP can enhance our measurement of $G$. In addition, we derive the relationship between EP enlarged eigenfrequency shift and the Newtonian constant. This work provides a way to engineer EP-assisted optomechanical devices for applications in the field of precision measurement of $G$
\end{abstract}

\maketitle


\section{\label{sec:level1}Introduction}
Newton's law of gravitation is often written as
\begin{equation}
	F=G\frac{m_1 m_2}{r^2},
\end{equation}
where $ m_1$ and $m_2$ are the masses of two particles, $r$ is the distance between them and $G$ is the gravitational constant. Though the absolute value of $G$ has been measured by many experiments \cite{Xue2020Precision,jianpingLiu2018,gillies1987newtonian}, two big puzzles still exist nowadays. One is the value of G ramains the least preciously known of the fundamental constants \cite{li2018measurements}. The Committee on Data for Science and Technology (CODATA) recommended value of $G$ based on the 2014 adjustment is $6.67408(31)\times {10}^{-11} m^3 {kg}^{-1} s^{-2}$, and the relative measurement uncertainty is as high as $ 4.7\times {10}^{-5}$ \cite{mohr2016rev}. The other puzzle is the experimental values of $G$ are not consistent with each other\cite{2019On}. To resolve these puzzles strongly motivates us to develop different measurement methods \cite{quinn2014don}.

In this paper, we develop a new method for determining $G$. In our method, a  quantum system consisting of two cavities and two membranes is considered. Gravitational force gradient originating from the source mass causes shifts of the resonant frequencies of two membranes, resulting the shifts of the eigenfrequencies of the supermodes emerging in our system. Based on this, we establish our measurement principles. Exceptional points are branch point singularities of the spectrum and eigenfunctions \cite{Heiss_2004} which occur generically in eigenvalue problems that depend on a parameter\cite{2012The}. Because EPs in open quantum and wave systems can exhibit a strong spectral response to perturbations\cite{Wiersig:20}, their potential applications in precision measurement have been investigated both theoretically\cite{PhysRevA.93.033809,PhysRevLett.112.203901} and experimentally\cite{hodaei2017enhanced,Chen2017Exceptional}. In this paper, EP effect is investigated numerically. The relating results indicate that this effect can enhance the detection of the eigenfrequency shift. Furthermore, the relationship between  eigenfrequency shift and Newtonian constant is established  provided that our system is operated at EP. In summary, we propose an EP-based quantum mechanical method for measuring $G$.  Finally we expect our work can enrich the experimental approaches of determining $G$.

The remainder of the paper is organized as follows: In Sec. II we demonstrate the theoretical framework, in Sec. III we present the measurement principles,
in Sec. IV we summarize the paper and provide an outlook.
\begin{figure}[b]
	\includegraphics[width=25em]{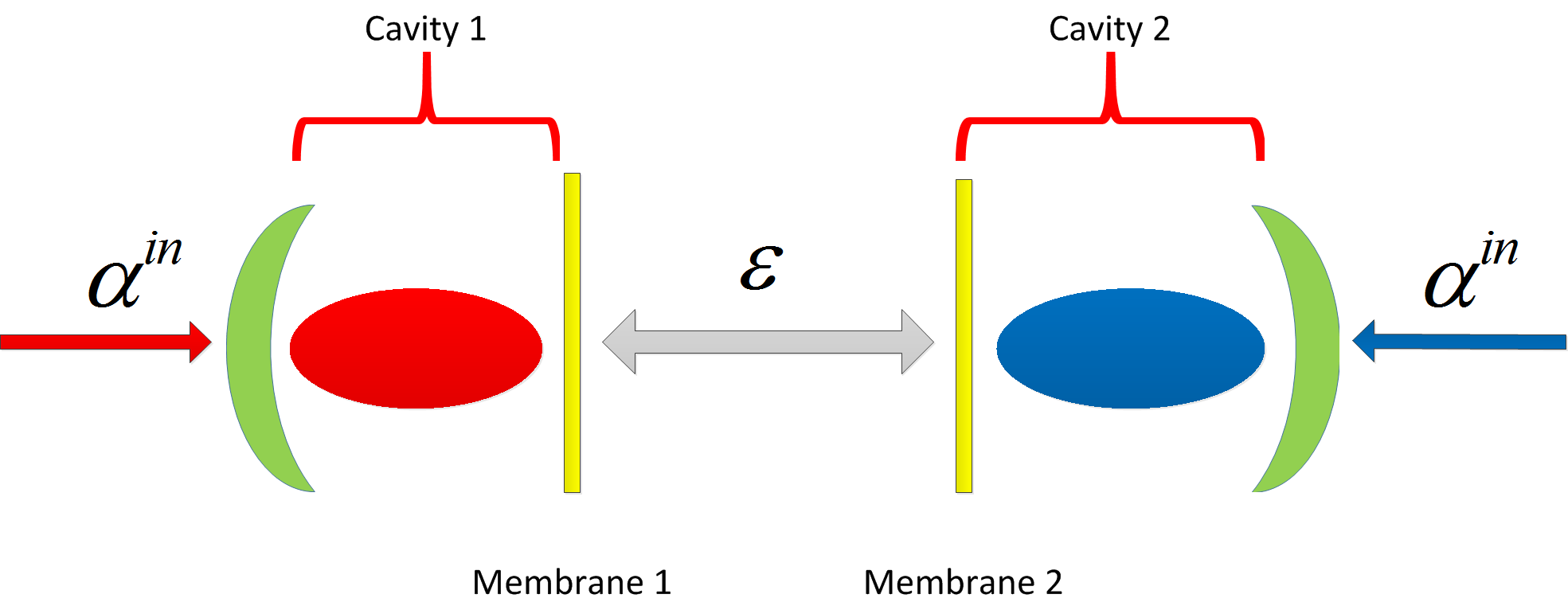}
	\caption{\label{fig:epsart} Schematic setup. Two Cavities, whose end-mirrors are two membrane resonators, are driven with two detuned lasers. }
\end{figure}
\section{\label{sec:level1}Theoretical framework}

We consider a system composed of two optomechanical cavities the two end-mirrors of which are two nanosized membrane resonators \cite{aspelmeyer2014cavity}.  Cavity 1(2) is driven with a red (blue) detuned laser. By symmetrically driving the cavities we can engineer either mechanical gain or mechanical loss \cite{djorwe2019exceptional}. Note that at the gain and loss balance the system features EP. In the rotating frame of the driving fields, the Hamiltonian of the system could be written as ($\hbar=1$)
\begin{equation}
	H=H_1 +H_2 +H_3,
\end{equation}
where \begin{align}
	H_1&=\sum_{j=1,2} \omega_j b_j^+ b_j -\Delta_j a_j^+ a_j -g a_j^+ a_j (b_j^+ + b_j)\notag\\
	H_2&=-\varepsilon(b_1 b_2^+ +b_1^+ b_2)\notag\\
	H_3&=\sum_{j=1,2}E(a_j^+ +a_j).
\end{align}
Here $a_j$ and $b_j$ are the annihilation bosonic field operators describing the optical and mechanical resonators. $\omega_j$ is the mechanical frequency of the $j^{th}$ resonator and $\Delta_j =\omega_P^j -\omega_c^j$ is the optical detuning between the optical ($\omega_P^j$) and the cavity ($\omega_c^j$) frequencies. $\varepsilon$ is the coupling strength between the two mechanical resonators, $g$ is the vacuum optomechanical coupling strength. The amplitude of the driving pump is $E$.

The quantum Langevin equations (QLEs) for the operations of the optical and the mechanical modes are derived from Eq. (3) as
\begin{align}
{\dot a}_j&=[i(\Delta_j+g(b_j^+ + b_j))-\frac{\kappa}2] a_j-i\sqrt{\kappa}(\alpha^{in}+\xi_{a_j})\notag\\
{\dot b}_j&=-(i\omega_j+\frac{\gamma_m}2)b_j+i\varepsilon b_k+iga_j^+a_j +\sqrt{\gamma_m} \xi_{b_j},
\end {align}
where optical ($\kappa$) and mechanical ($\gamma_m$) dissipations have been added,  $k$ is defined as $k=3-j$,
and the amplitude of the driving pump has been substituted as $E=\sqrt{\gamma_m} \alpha^{in}$ in order to account for losses. The term $\xi_{a_j}$ ($\xi_{b_j}$) denotes the optical (thermal) Langevin noise at room temperature.
We seek to investigate in the classical limit, where photon and phonon numbers are assumed large in the system, and noise terms can be neglected in our analysis. Then we can derive the following set of nonlinear equations:
\begin{align}
{\dot \alpha}_j&=[i(\Delta_j+g(\beta_j^* + \beta_j))-\frac{\kappa}2] \alpha_j-i\sqrt{\kappa} \alpha^{in}\notag\\
{\dot \beta}_j&=-(i\omega_j+\frac{\gamma_m}2)\beta_j +i\varepsilon \beta_k+ig\alpha_j^* \alpha_j.
\end {align}
Here $\alpha_j =\langle a_j \rangle $ and $\beta_j =\langle b_j \rangle$.

To identify the EP feature, we approach the limit cycle oscillations by the ansatz, $\beta_j (t)={\bar\beta}_j +A_j exp(-i\omega_{lock} t)$. ${\bar\beta}_j$ is a constant shift in the origin of the movement, $A_j$ is taken to be a slowly varying function of time, and $\omega_{lock}$ is the mechanical locked frequency. According to \cite{djorwe2018frequency}, we can derive
\begin{align}
i\frac{\partial\Psi}{\partial t}=H_{eff}\Psi,
\end{align}
where $\Psi=(\beta_1, \beta_2)^T$ is the state vector and the effective Hamiltonian is
		
\[H_{eff}=\begin{bmatrix}
\omega^1_{eff}-i{\frac{\gamma^1_{eff}}2}&-\varepsilon\\-\varepsilon&\omega^2_{eff}-i{\frac{\gamma^2_{eff}}2}
\end{bmatrix} \]
. Here $\omega^j_{eff}=\omega_j + \Delta \omega_j$ and $\gamma^j_{eff}=\gamma_m +\gamma_{opt}^j$ define the effective frequencies and the effective damping respectively. The optical spring effect $\Delta \omega_j$ and the optical damping $\gamma_{opt}^j$ are given by
\begin{align}
\Delta \omega_j=-\frac{2\kappa(g \alpha^{in})^2}{\omega_{lock} \zeta_j} Re(\sum_n \frac{J_{n+1}(-\zeta_j)J_n (-\zeta_j)}{h_{n+1}^{j*} {h_n^j}})
\end{align}
and
\begin{align}
\gamma_{opt}^j=\frac{2(\kappa g \alpha^{in})^2}{ \zeta_j} \sum_n \frac{J_{n+1}(-\zeta_j)J_n (-\zeta_j)}{{|h_{n+1}^{j*} {h_n^j}|}^2}.
\end{align}
$\zeta_j=\frac{2g Re(A_j)}{\omega_{lock}}$,$ {\tilde\Delta}_j=\Delta_j+2gRe({\bar\beta}_j)$, $h_n^j=i(n\omega_{lock}-{\tilde\Delta}_j)+\frac{\kappa}2$, and $J_n$ is the Bessel function of the first kind of order  $n$. The eigenvalues of the effective Hamiltonian can be described as
\begin{align}
\tau_{\pm}=\frac{\omega^1_{eff}+\omega^2_{eff}}2-i\frac{\gamma^1_{eff}+\gamma^2_{eff}}4 \pm\frac{\sqrt\varDelta}2,
\end{align}
with
\begin{align}
\varDelta=[(\omega^1_{eff}-\omega^2_{eff})+\frac i2 (\gamma^2_{eff}-\gamma^1_{eff})]^2 +4\varepsilon^2.
\end{align}
The EP of our system appears if $\Delta=0$.
The eigenfrequencies and the dissipations of the supermodes are defined as the real and imaginary parts of the eigenvalues, i.e.,$\nu_{\pm}=Re(\tau_\pm)$, and $\Upsilon_{\pm}=Im(\tau_\pm)$. At the EP, both pairs of eigenfrequencies and dissipations coalesce, i.e., $\nu_+ =\nu_-$ and $\Upsilon_+ =\Upsilon_-$.

For simplicity, we assume the two resonators are degenerated, causing $\omega_1=\omega_2=\omega_r$. We adopt a  weak driving regime, resulting $\Delta \omega_j\ll \omega_r$.  We further assume $\gamma_{opt}^j=(\alpha^{in})^2 \eta_j$ with $\eta_1 \neq\eta_2$.
Then the eigenvalues of the Hamiltonian can be written as
\begin{align}
\tau_{\pm}\approx &\omega_r -i\frac{2\gamma_m +(\eta_1 +\eta_2) (\alpha^{in})^2}4 \notag\\ &\pm\sqrt{\varepsilon^2-\frac{(\alpha^{in})^4 (\eta_2 -\eta_1)^2}{16}}.
\end{align}
As a result, the EP condition is tranformed into $\varepsilon^2-\frac{(\alpha^{in})^4 (\eta_2 -\eta_1)^2}{16}=0$, from which we derive that $\alpha^{in}=\alpha_{EP}^{in}=\sqrt\frac{4\varepsilon}{|\eta_2 -\eta_1|}$. Now we use an example to illustrate the EP feature (see Fig. 2). In this example, the parameters used are $\eta_1={10}^{-6}, \eta_1=2\times{10}^{-6}, \varepsilon={10}^{-2}\omega_r, \gamma_m ={10}^{-4}\omega_r.$ The corresponding EP is $\alpha_{EP}^{in}=200\omega_r^{1/2}$. From the figure, it is seen that two eigenfrequencies coalesce at the EP as well as two dissipations.
Based on the above analysis about our system, we present the principles of measurement in the next section.
\begin{figure}[b]
\includegraphics[width=25em]{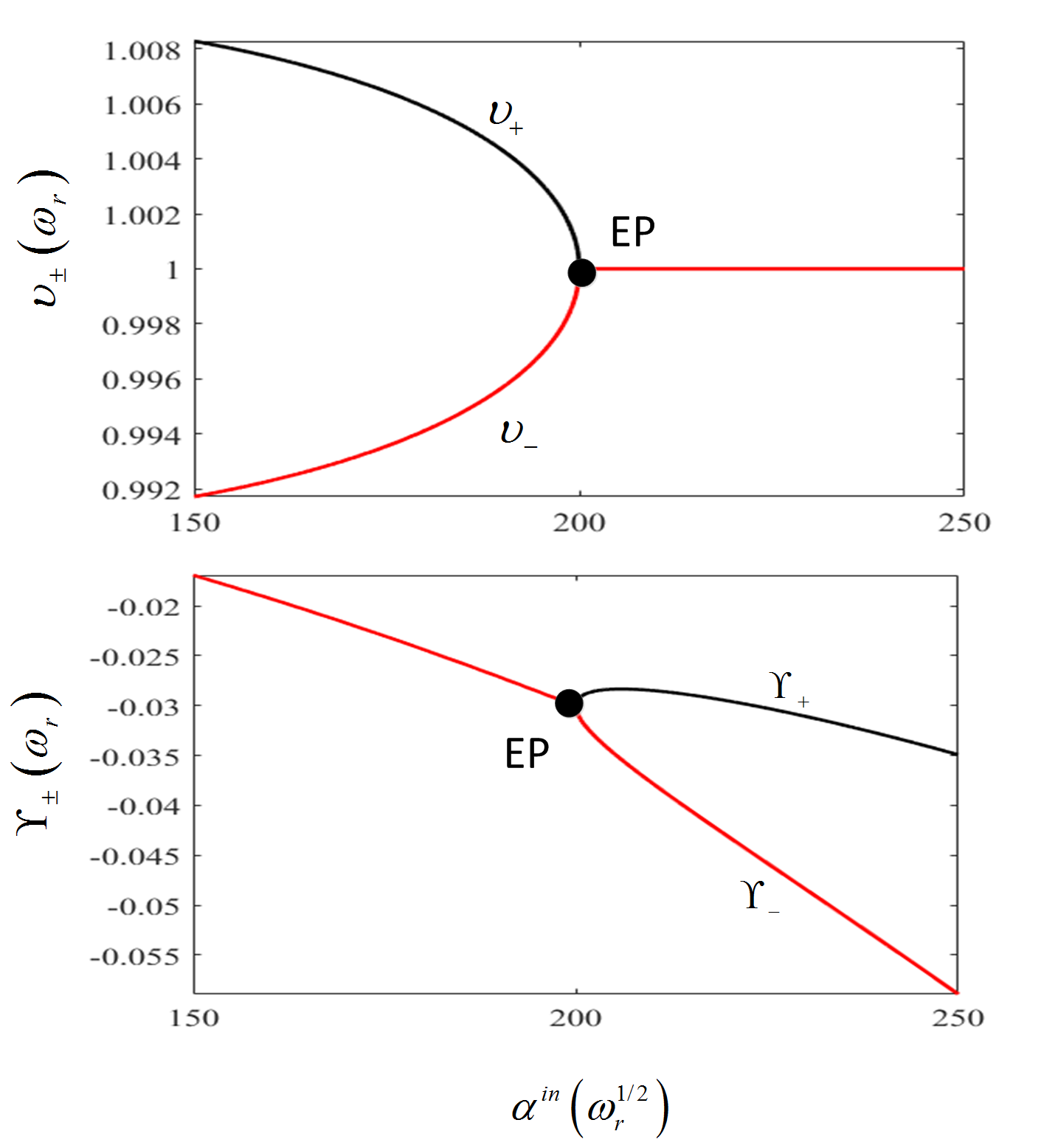}
\caption{\label{fig:epsart} Illustration of EP feature. Eigenfrequencies $\nu_{\pm}$ and dissipations $\Upsilon_{\pm}$ versus the driving strength $\alpha ^{in}$. the parameters used are $\eta_1={10}^{-6}, \eta_1=2\times{10}^{-6}, \varepsilon={10}^{-2}\omega_r, \gamma_m ={10}^{-4}\omega_r.$ }
\end{figure}

\section{\label{sec:level1}Principles of measurement}
\begin{figure}[b]
	\includegraphics[width=25em]{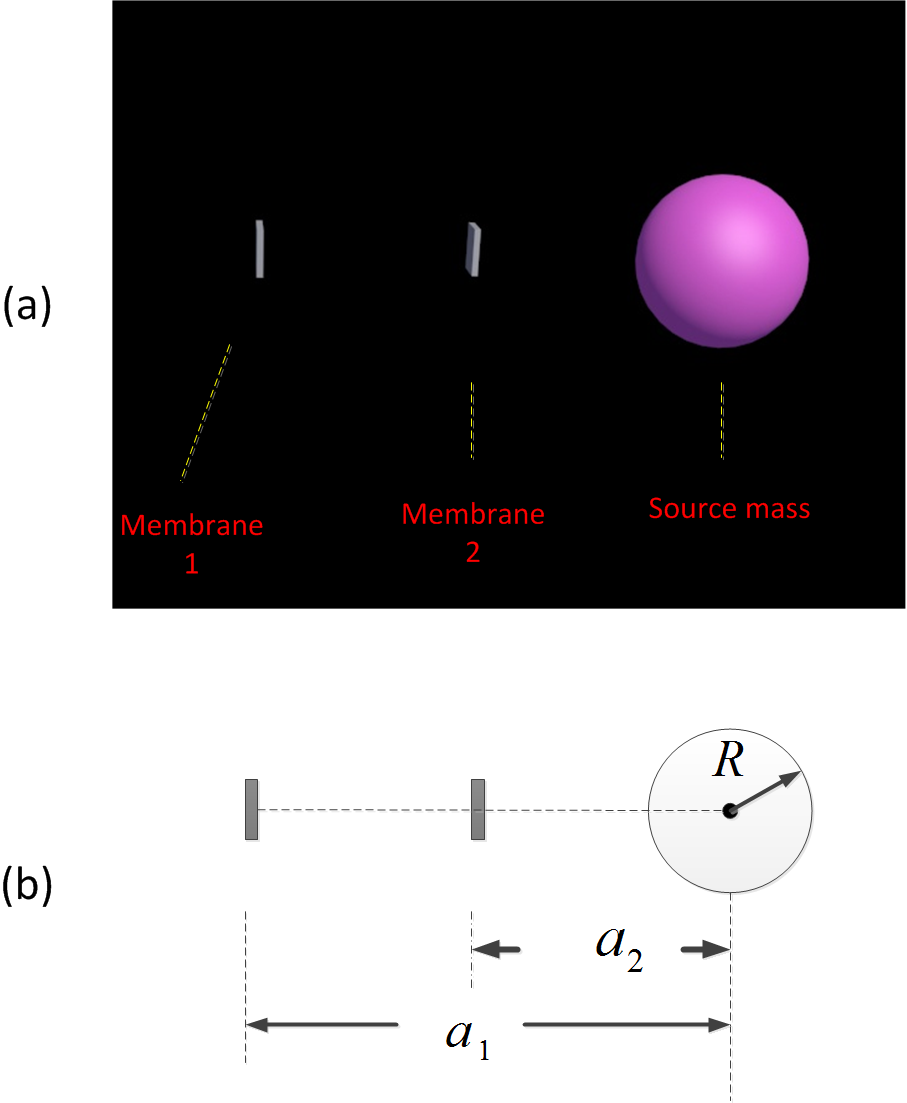}
	\caption{\label{fig:epsart} (a)The locations of two membranes and the sphere . (b) A cross section of the diagram in (a). }
\end{figure}
In our proposed method, a sphere with 
uniformly distributed mass $m$, radius $R$ and density $\rho$ is utilized. This sphere and two membranes of our system are placed as shown in Fig. 3(a)
Note that the sizes of two membrane resonators are much smaller than $R$. The distances between two membranes and the center of the sphere are  $a_1$ and $a_2$  respectively, which is shown in Fig. 3(b). We assume that exotic forces such as non-Newtonian gravitylike forces applied to the two membranes can be neglected. Because of the gravitational forces generated by the source mass, the resonant frequencies of two resonators are modified by $\delta \omega_1$ and $\delta \omega_2$ respectively. Here $\delta \omega_1$ and $\delta \omega_2$ are defined as
$\delta \omega_j =\omega_j^{\prime}-\omega_j$, where $\omega_j^{\prime}$ is the perturbed resonant frequency. From \cite{PhysRevA.72.033610} we obtain
\begin{equation}
	\frac{\delta \omega_j}{\omega_j}\approx \frac 1{2m_j \omega_j^2} \frac{\partial F(a_j)}{\partial a_j},
\end{equation}
where $m_j$ is the mass of $j$th membrane,  and the gravitational force acting on $j$th membrane is
\begin{equation}
	F(a_j)=\frac{GM m_j}{a_j^2}.
\end{equation}
From the above two equations, we derive
\begin{equation}
	\delta \omega_j \approx -\frac{GM}{\omega_j a_j^3}.
\end{equation}
From section 2 we know $\omega_1=\omega_2=\omega_r$. Then  we assume that $a_1 \gg a_2$, resulting $|\delta \omega_1|\ll |\delta\omega_2|$.
\begin{figure*}
	\includegraphics[width=50em]{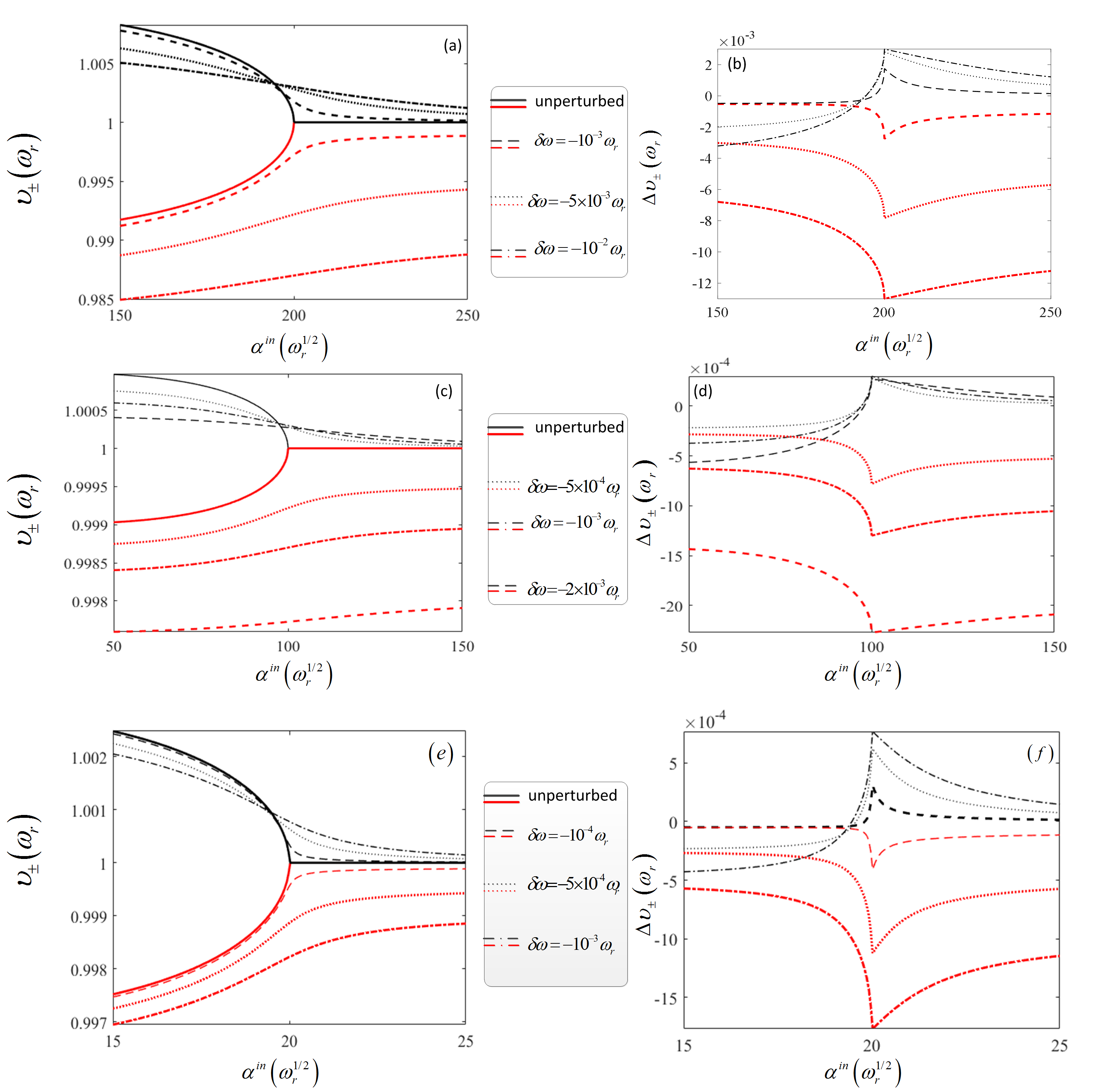}
	\caption{\label{fig:epsart} Eigenfrequencies $\nu_{\pm}$ and eigenfrequencies shifts $\Delta\nu_{\pm}$ versus the driving field $\alpha_{in}$ around EP. The black curves are for $\nu_{+}$ and $\Delta\nu_{+}$, while red ones for $\nu_{-}$ and $\Delta\nu_{-}$. (a)-(b)The parameters used are $\eta_{1X} ={10}^{-6}, \eta_{2X} =2\times{10}^{-6}, \varepsilon_X={10}^{-2}\omega_r, \gamma_{mX}={10}^{-4}\omega_r.$ (c)-(d)The parameters used are $\eta_{1Y} =3\times{10}^{-7}, \eta_{2Y} =7\times{10}^{-7}, \varepsilon_Y={10}^{-3}\omega_r, \gamma_{mY}={10}^{-3}\omega_r$. (e)-(f) The parameters used are $\eta_{1Z} ={10}^{-5}, \eta_{2Z} =4\times{10}^{-5}, \varepsilon_Z=3\times{10}^{-3}\omega_r, \gamma_{mZ}=2\times{10}^{-3}\omega_r.$ }
\end{figure*}

As a result of resonant frequency modification of two membranes, the eigenvalues of the Hamiltonian can be rewritten as
\begin{align}
	\tau_{\pm}^{\prime} \approx & \omega_r + \frac{\delta \omega_2}2
	-i\frac{2\gamma_m + (\alpha^{in})^2(\eta_1+\eta_2)}4\notag\\ &\pm\frac{\sqrt{[-\delta \omega_2 +\frac i2 (\alpha^{in})^2 (\eta_2-\eta_1)]^2+4\varepsilon^2}}2
\end{align}
For convenience, we substitute $\delta\omega$ for $\delta\omega_2$ in the remaining section. The shifts of the eigenfrequencies of the supermodes caused by the source mass can be described as
\begin{equation}
	\Delta\nu_{\pm}=Re(\tau_\pm^\prime)-Re(\tau_\pm).
\end{equation}
Then from Eqs. (11), (15)-(16) we  derive that
\begin{align}
	&\Delta\nu_\pm =Re(\frac{\delta\omega}2 \notag\\&\pm\sqrt{\varepsilon^2-\frac{(\alpha^{in})^4 (\eta_2-\eta_1)^2}{16}+\frac{(\delta\omega)^2-i\delta\omega(\alpha^{in})^2 (\eta_2-\eta_1)}4}\notag\\&\mp\sqrt{\varepsilon^2-\frac{(\alpha^{in})^4 (\eta_2-\eta_1)^2}{16}}
\end{align}
In the following, we consider three cases denoted by X, Y, and Z where we operate our system near the EP.

In the case of X, the parameters used are $\eta_{1X} ={10}^{-6}, \eta_{2X} =2\times{10}^{-6}, \varepsilon_X={10}^{-2}\omega_r, \gamma_{mX}={10}^{-4}\omega_r,$ and the corresponding EP  is $\alpha_{EPX}^{in}=200\omega_r^{1/2}$. In Y, $\eta_{1Y} =3\times{10}^{-7}, \eta_{2Y} =7\times{10}^{-7}, \varepsilon_Y={10}^{-3}\omega_r, \gamma_{mY}={10}^{-3}\omega_r, \alpha_{EPY}^{in}=100\omega_r^{1/2}.$ In Z, $\eta_{1Z} ={10}^{-5}, \eta_{2Z} =4\times{10}^{-5}, \varepsilon_Z=3\times{10}^{-3}\omega_r, \gamma_{mZ}=2\times{10}^{-3}\omega_r, \alpha_{EPZ}^{in}=20\omega_r^{1/2}.$ Figure 4 illustrates how eigenfrequencies ($\nu_{\pm}$) and eigenfrequencies shifts ($\Delta\nu_{\pm}$) undergo the EP corresponding to different perturbation. (a)-(b), (c)-(d) and (e)-(f) are for X, Y, Z respectively. The black curves are for $\nu_{+}$ and $\Delta\nu_{+}$, while red ones for $\nu_{-}$ and $\Delta\nu_{-}$. From this figure, we find that in all nine situations (different perturbation and different three cases) we would obtain a relative bigger positive frequency shift ($|\Delta\nu_{+}|$) and a maximum negative eigenfrequency shift ($|\Delta\nu_{-}|$) at EPs.

\begin{figure}[b]
	\includegraphics[width=25em]{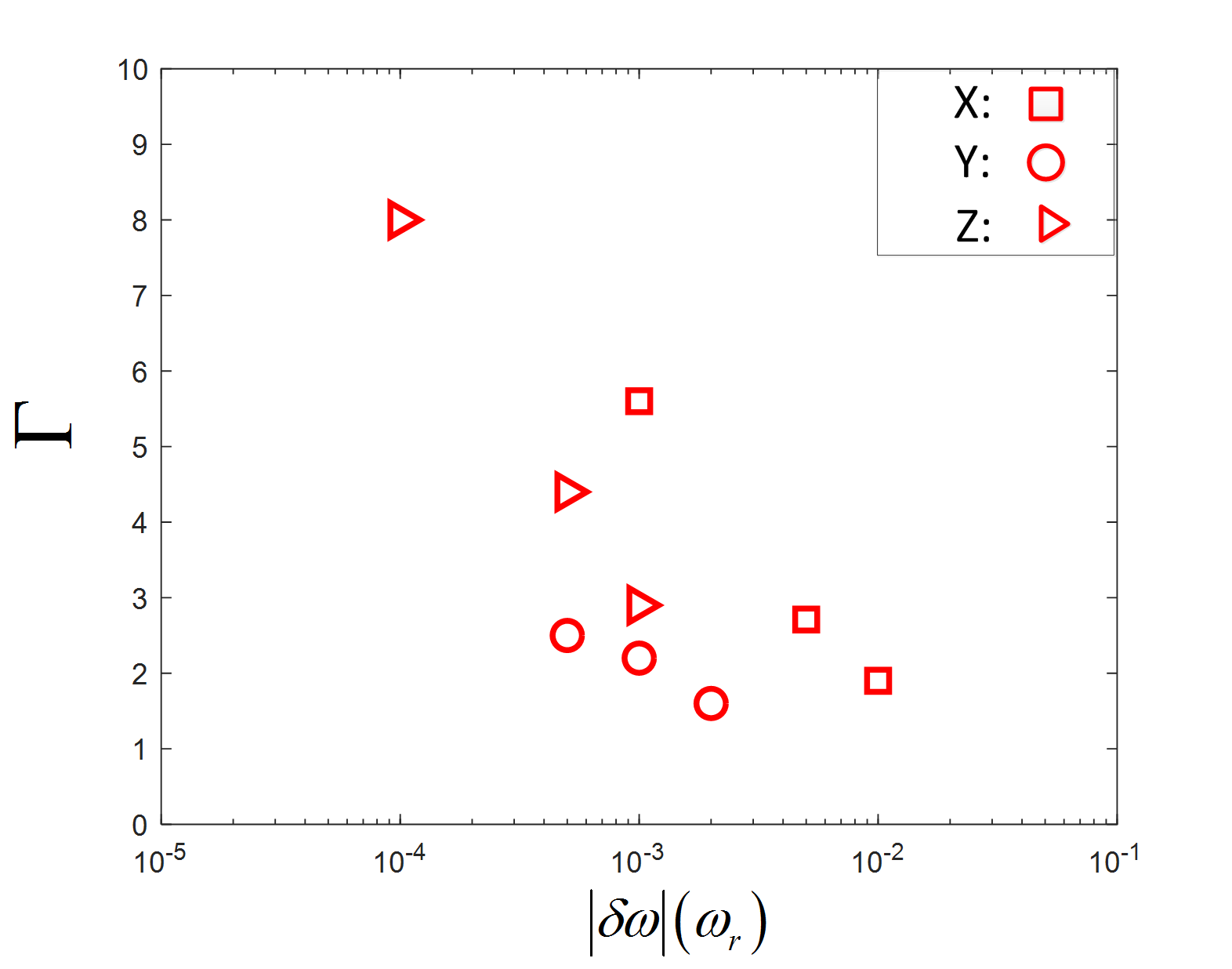}
	\caption{\label{fig:epsart} Nine ratios ($\Gamma$)correspond to three cases(X, Y, and Z) and different perturbation ($|\delta\omega|$).}
\end{figure}

To further demonstrate the EP effect, 
we define ratio $\Gamma$ as $\Gamma=\frac{|\Delta\nu_{-}|_{EP}}{|\Delta\nu_{-}|_{min}}$, where $|\Delta\nu_{-}|_{EP}$ is the value of $|\Delta\nu_{-}|$ related to $\alpha_{EP}^{in}$ and $ |\Delta\nu_{-}|_{min}$ is the minimum value of $|\Delta\nu_{-}|$. Nine values of $\Gamma$ correspond to three cases(X, Y, and Z) and different perturbation ($|\delta\omega|$) are shown in Fig. 5.
where square, circles and triangle correspond to X, Y and Z respectively.
We can find that in any of three cases a smaller $|\delta\omega|$ corresponds to a bigger $\Gamma$. Since $\delta\omega$ in practical experiments would be very tiny, we conclude that  operating the system at EP may make a large contribution to our measurement.

We assume our system is operated at EP and investigate the relationship between EP enlarged eigenfrequency shift and the Newtonian constant. Accordingly there is $\alpha^{in}=\alpha^{in}_{EP}=\sqrt{\frac{4\varepsilon}{|\eta_2 -\eta_1|}}$. With the assumption $\eta_2 >\eta_1$,
according to Eq. (17) we  derive that
\begin{equation}
	\Delta\nu_{\pm}=Re(\frac{\delta\omega}2 \pm\sqrt{\frac{(\delta\omega)^2 -i\delta\omega4\varepsilon}4}).
\end{equation}
Then, from Eq. (18) we derive
\begin{equation}
	\Delta\nu_{\pm}=\frac{\delta\omega}2 (1\mp\sqrt{\frac{1+\sqrt{1+\frac{16\varepsilon^2}{(\delta\omega)^2}}}2}).
\end{equation}
Next we focus on $\Delta\nu_{-}$. From the above, we obtain
\begin{equation}
	\delta\omega\approx-\frac{GM}{\omega_r a_2^3}.
\end{equation}
From Eqs. (19)-(20), we derive
\begin{align}
	\Delta\nu_{-}=-\frac{GM}{2\omega_r a_2^3}-\frac{GM}{2\omega_r a_2^3}\sqrt{\frac{1+\sqrt{1+\frac{16\varepsilon^2 \omega_r^2 a_2^6 }{G^2 M^2}}}2}.
\end{align}
We assume $a_2-R\ll R$. Consequently there is $a_2\approx R$. Since $M=\frac{4}{3} \pi R^3 \rho$, equation (21) can be rewritten as
\begin{equation}
	\Delta\nu_{-}\approx -\frac{2\pi G\rho}{3\omega_r}-\frac{2\pi G\rho}{3\omega_r}\sqrt{\frac{1+\sqrt{1+\frac{9\varepsilon^2 \omega_r^2  }{G^2 \pi^2 \rho^2}}}2}.
\end{equation}
Then we obtain
\begin{equation}
	|\Delta\nu_{-}|= \frac{2\pi G\rho}{3\omega_r}+\frac{2\pi G\rho}{3\omega_r}\sqrt{\frac{1+\sqrt{1+\frac{9\varepsilon^2 \omega_r^2  }{G^2 \pi^2 \rho^2}}}2}.
\end{equation}
Till now, we have derived the relationship between eigenfrequency shift and the Newtonian constant. Then we can determine the value of $G$ and the relative uncertainty via the detection of $|\Delta\nu_{-}|$.

\begin{figure}[b]
	\includegraphics[width=25em]{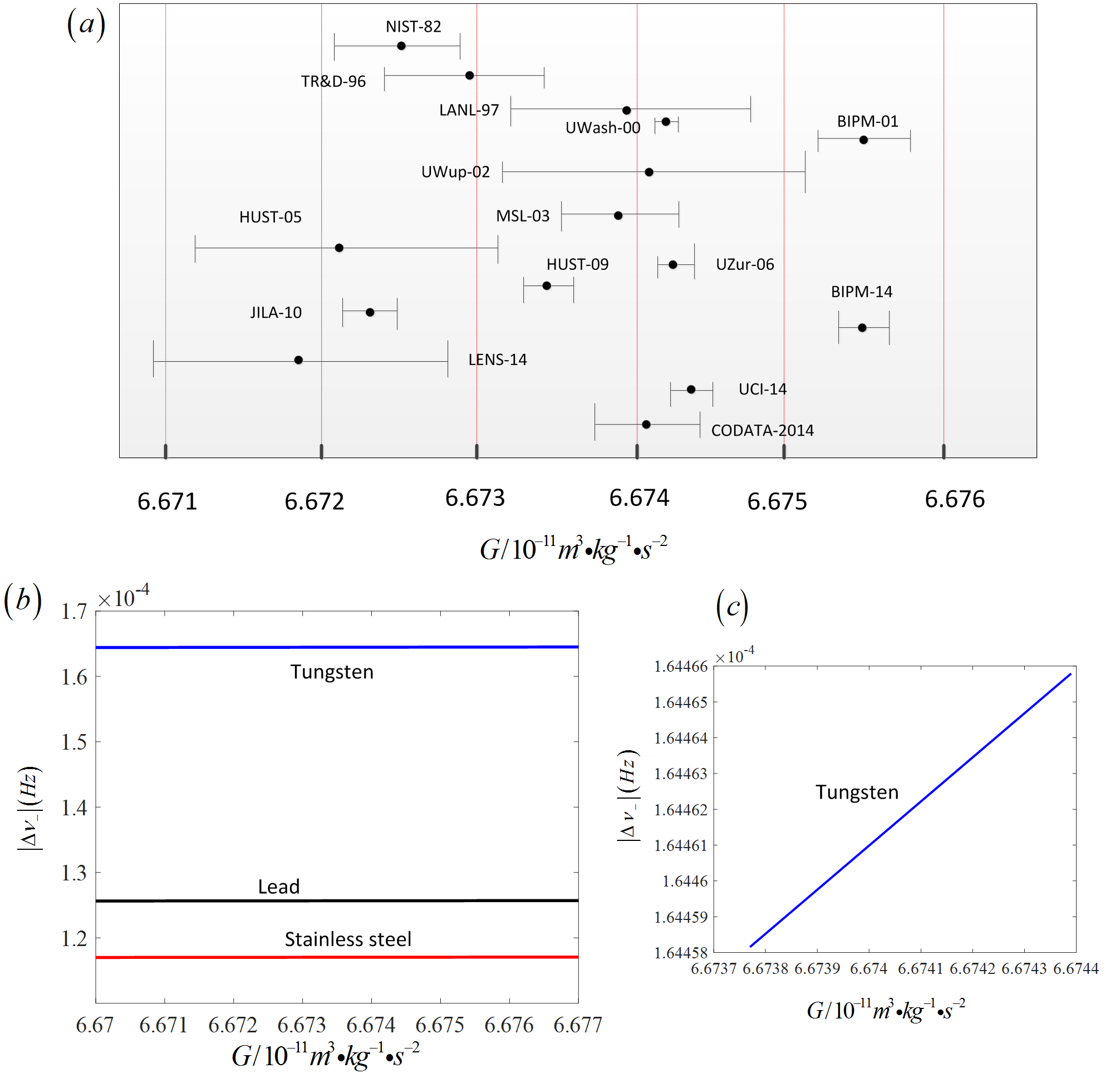}
	\caption{\label{fig:epsart} (a) The adopted values of G in CODATA-2014 adjustment. (b)-(c) $|\Delta\nu_{-}|$ as a function of $G$. The parameters used are $\varepsilon={10}^{-2}\omega_r, \omega_r=2\times{10}^{9}Hz$.  (b) deals with three spheres with different densities, which are made of stainless steel, lead, and tungsten respectively. On the contrary, (c) only deal with a Tungsten sphere. Note that in (c), 
		$G$ takes values of $6.67408(31)\times{10}^{-11} m^3\bullet {kg}^{-1} \bullet s^{-2}$.}
\end{figure}

In Fig.6(a), the adopted values of $G$ in CODATA-2014 adjustment \cite{mohr2016rev,luther1982redetermination,karagioz1996measurement,bagley1997preliminary,gundlach2000measurement,quinn2001new,kleinevoss2002bestimmung,armstrong2003new,hu2005correction,holzschuh2006measurement,luo2009determination,tu2010new,parks2010simple,quinn2013improved,rosi2014precision,newman2014measurement} are illustrated according to \cite{jianpingLiu2018}. To visualize our result (Eq.(23)), we plot $|\Delta\nu_{-}|$ as a function of $G$ in Fig.6(b)-(c), where
the parameters used are $\varepsilon={10}^{-2}\omega_r$, and $\omega_r=2\times{10}^{9}Hz$.
In (b), we utilize a Stainless steel sphere with $\rho = 9.8\times {10}^3 Kg/m^3 $,  a Lead one with 
$\rho = 11.3 \times {10}^3 Kg/m^3 $ and a Tungsten one with 
$\rho = 19.35 \times {10}^3 Kg/m^3 $.
On the contrary, only a Tungsten sphere is considered in (c), where
$G$ takes values of $6.67408(31)\times{10}^{-11} m^3\bullet {kg}^{-1} \bullet s^{-2}$  which is the 2014 CODATA recommended value of $G$.

\section{\label{sec:level1}Summary and outlook}
In sum, this paper presents a novel method which can be considered in the  field of precision measurement of $G$. In this method, two cavities and two membrane resonators constitute an optomechanical system. Gravitational force gradient perturbs the resonant frequencies of two membranes, resulting the shifts of the eigenfrequencies of the supermodes.  Based on EP enhanced  shifts, we can perform the measurement of  $G$. Compared to traditional measurements methods \cite{rothleitner2017invited,rosi2016challenging} such as torsion balance, atom interferometry, etc., our method possesses two distinct characters. They are: a. The proposed setup is a minute-sized optomechanical system. b. EP effect plays an important role in the measurement of $G$.

Based on this work where a second-order EP feature emerges in the proposed system,  we can attempt to construct a  new system which can be operated at higher-order EP, and a higher sensitivity may be achieved \cite{jing2017high,hodaei2017enhanced,miri2019exceptional}. So far, EP effect in optomechanical systems have been observed \cite{Jing2018A,Harris2016Topological}, indicating that our method can be put into consideration. Moreover, \cite{2014Parity} can be referred to for the detection of the shift of the supermode eigenfrequency. Finally, we expect our proposal can promote the application of EP-based sensors to the measurement of $G$.

\begin{acknowledgments}
This work was supported by Natural Science Foundation of Shanghai (No. 20ZR1429900).
\end{acknowledgments}

\bibliography{apssamp}

\end{document}